# calibrar: an R package for fitting complex ecological models

Ricardo Oliveros-Ramos[1,2,3] and Yunne-Jai Shin[2,3]

[1]Instituto del Mar del Perú (IMARPE). Gamarra y General Valle s/n Chucuito, Callao, Perú.

[2]MARBEC, Institut de Recherche pour le Développement (IRD), Univ Montpellier, IFREMER, CNRS, Montpellier, France.

[3]University of Cape Town, Marine Research (MA-RE) Institute, Department of Biological Sciences, Private Bag X3, Rondebosch 7701, South Africa.

Corresponding author: Ricardo Oliveros-Ramos (ricardo.oliveros@ird.fr)

**Summary**

1. The fitting or parameter estimation of complex ecological models is a challenging optimisation task, with a notable lack of tools for fitting complex, long runtime or stochastic models.

2. calibrar is an R package that is dedicated to the fitting of complex models to data. It is a generic tool that can be used for any type of model, especially those with non-differentiable objective functions and long runtime, including Individual Based Models.

3. calibrar supports multiple phases and constrained optimisation, includes 18 optimisation algorithms, including derivative-based and heuristic ones. It supports any type of parallelization, the restart of interrupted optimisations for long runtime models and the combination of different optimisation methods during the multiple phases of the calibration.

4. User-level expertise in R is necessary to handle calibration experiments with calibrar, but there is no need to modify the model's code, which can be programmed in any language. It implements maximum likelihood estimation methods and automated construction of the objective function from simulated model outputs. For more experienced users, calibrar allows the implementation of user-defined objective functions.

5. The package source code is fully accessible and can be installed directly from CRAN.

## Introduction

The ability to achieve accurate parameter estimation has been used as a criterion to assess the usefulness of ecological models (Bartell 2003). Given a model, the criterion for the selection of the best possible parameter set is the optimisation of a scalar *objective function* (e.g. log-likelihood, residual sum of squares) with respect to the model parameters (Walter and Pronzato 1997, Bolker et al. 2013). Once the objective function is properly defined, parameter estimation is essentially an optimisation problem. Parameter estimation, calibration or fitting of ecological models (Jorgensen and Bendoricchio 2001) can be a challenging task for optimisation algorithms when dealing with complex models characterized by non-linearity and high dimensionality, as well as low quantity and quality of observed data (Tashkova et al. 2012). To date, there has been limited development of optimisation algorithms and calibration methodologies for complex ecological models that are sufficiently flexible, generic, and well-documented for users (Bolker et al. 2013). Additionally, complex ecosystem models can be numerically intensive and require long simulation runs, adding an extra layer of difficulty to their fitting process.

There are some dedicated tools for non-linear parameter estimation, AD Model Builder (ADMB; Fournier et al. 2012) being one of the most robust and fast (Bolker et al. 2013). Among other advantages, ADMB provides support for parameter estimation in multiple phases or "masks", with some parameters temporarily fixed (Nash and Walker-Smith 1987), which can be of great interest when dealing with complex ecosystem models (Oliveros-Ramos et al. 2017). It also provides support for bound or box constraints (independent fixed bounds on each parameter), which can be helpful for regularizing hard optimisation problems (Bolker et al. 2013). However, the model and the objective function itself need to be coded in C++ (using a scripting language), which can be an obstacle for fitting models which have been already implemented in other languages (e.g. Fortran, Java). In addition, it is based in automatic differentiation (AD), which allows to provide accurate estimates of derivatives (Griewank and Corliss 1992), but does not handle the estimation of parameters of stochastic models for which derivatives cannot be computed.

Parameter estimation methods have been developed for stochastic non-linear models, such as continuous time, finite state Markov models and individual-based models (IBMs), for which the probability of state transitions or the master equation can be written (Ionides et al. 2006, Newman et al. 2009, Ross et al. 2009, Walker et al. 2006). However, many stochastic models at the individual level can only be simulated numerically and are too complex for mathematical analysis and explicit parameter estimation (Black and McKane 2012), resulting in more attention being given to the exploration of model behaviour than to a rigorous confrontation with data. To solve these issues, meta-heuristic algorithms (e.g. evolutionary algorithms) have been used (Cropper and Anderson 2004, Poovathingal and Gunawan 2010, Duboz et at. 2010, Tashkova et al. 2012, Travers-Trolet et al. 2013), and have in some cases shown better performance than derivative-based optimisation methods (Tashkova et al. 2012). However, the scientific community lacks generic, open and flexible enough tools for the parameter estimation of different types of models with different degrees of complexity and computational power requirements.

Here we present a novel R package, calibrar, designed for parameter estimation for a wide range of ecological models, including complex and stochastic models. The package combines various optimization functionalities in a single interface, enabling the implementation of the latest advancements in complex model calibration. The package provides support for multiple phases and box constrained optimisation with the possibility of using several algorithms available in R. In particular, by using a "black-box" approach, the package allows the calibration of models implemented in any programming language. It provides a generic interface with models and allows the construction of the objective function, within R, without requiring any changes in the models' code. Parallel support for computationally intensive models is also provided, and can be used with high performance computing systems in a simple manner, including the capability to restart an unfinished optimisation for models with a long runtime.

## General description of the package

The `calibrar` package is written in R (R Development Core Team 2023), and can be installed from CRAN. The development version is available at GitHub (https://github.com/roliveros-ramos/calibrar). The package was primarily designed to handle the parameter estimation of complex models, e.g. using a maximum likelihood approach. However, it can be used for fitting any model, and can also be used with user-defined objective functions. The package was designed for the optimisation of "black-box" functions (Jones et al. 1998), where analytical information about the function to be optimized and the model source code are assumed to be unavailable or impractical to modify (Rios and Sahinidis 2013). Our approach is hence "non-intrusive", making the model interact with the optimizer, i.e. the `calibrar` package, in two ways: i) receiving a set of parameters to run the model, and ii) providing the model outputs to be confronted with observed data. The package also helps in the construction of the objective function to be optimized in order to estimate model's parameters (Figure 1).

The package was designed in a way that minimal expertise in R is necessary to handle the model fitting and the main functionalities are embedded in three functions: `calibrate`, `calibration_data` and `calibration_ObjFn` (see Table 1). The user intervention is mainly required in the construction of the function to run the model and to retrieve the simulation outputs within R (`run_model` function, Figure 1). However, given R's flexibility and features for data manipulation, it is rather straightforward to develop such a function. Some complex ecological models have dedicated packages oriented to the analysis and simulation of their outputs that could be used to link with `calibrar`, e.g., RNetLogo (Thiel et al. 2012) for IBMs implemented in netLogo or osmose (www.osmose-model.org) for the OSMOSE ecosystem model (Shin and Cury 2001, 2004). The main function of the package is `calibrate`, which performs minimization of the objective function and returns the optimal parameters. It has a similar syntax as the base R optimisation function `optim` (see Table 2), and is also similar to most optimisation related functions in other R packages. Additionally, the `calibration_data` and `calibration_ObjFn` are provided to simplify the organization of the observed data and construct the objective function for the calibration, respectively.

Table 1. Functions of the `calibrar` package.

| Function | Returned objects | Description |
|---|---|---|
| `calibrate` | An object summarizing the calibration results | Performs a sequential parameter estimation of a model using multiple phases |
| `calibration_data` | A list of the observed values (the data) | Create a list of the observed data with the information provided by its main argument |
| `calibration_objFn` | A function, integrating the run of the model and the comparison of outputs with observed data | Create a new function, to be used as the objective function in the calibration. |
| `calibration_setup` | A data.frame with the necessary information to create the objective function using `calibration_objFn` | A wrapper for `read.csv` checking column names and data types for the table with the calibration information |
| `coef, summary, predict` | | R S3 methods for visualizing the results of the calibration |

In order to create the objective function, the users need to specify some information about the model outputs used for the calibration and how to combine them (Figure 1). More experienced users can create the objective function by directly integrating the run of the model and its comparison with observed data. The details for the creation of the objective function are explained in the next section.

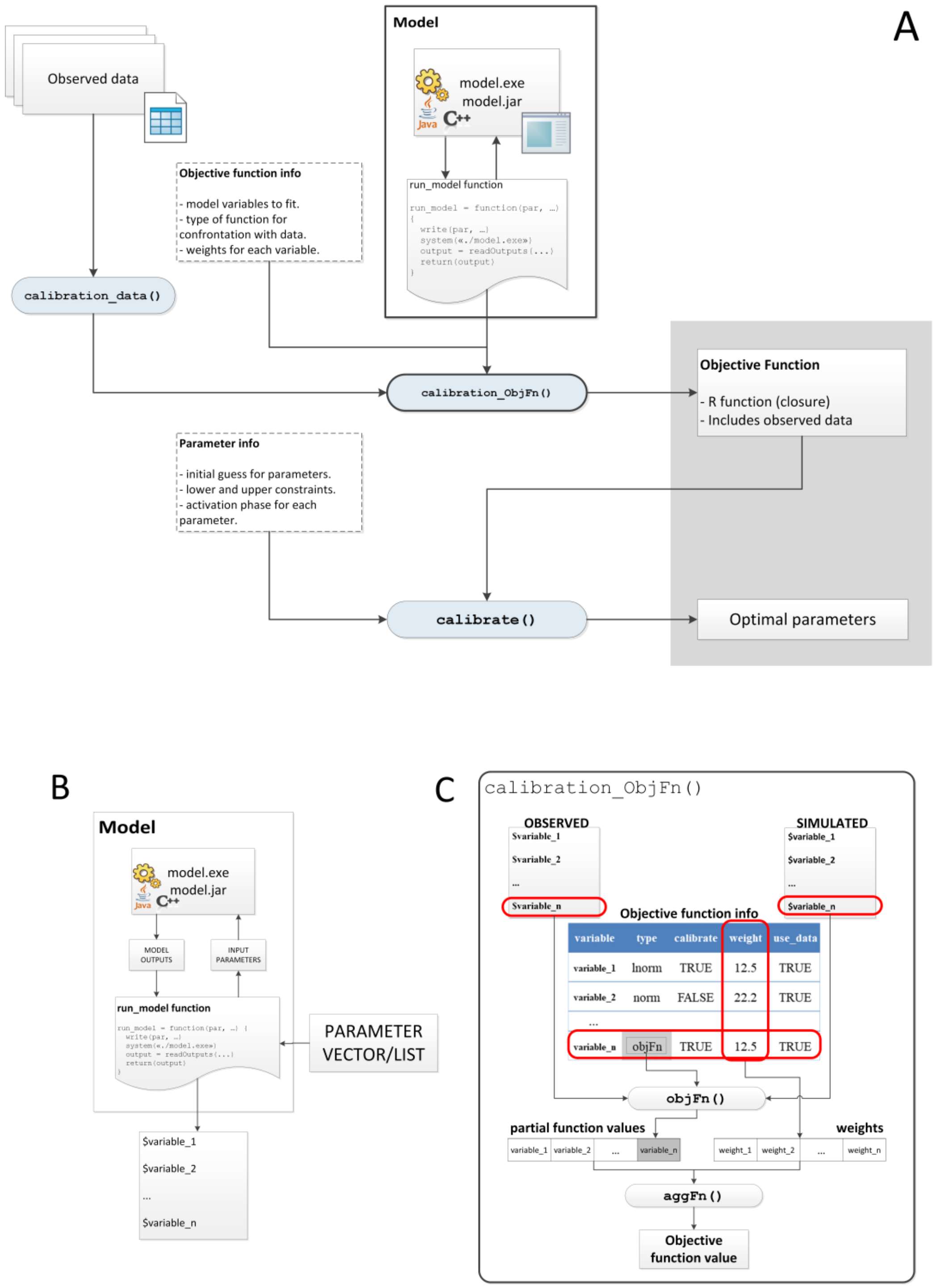

Figure 1. (A) Diagram representing the functioning of the `calibrar` package. The grey area groups the outputs produced by the package (the objective function and the optimal parameters of the model). Rectangles with broken border lines show user inputs which are needed to configure the model fitting process. Blue rounded rectangles show main package functions. (B) Scheme of the link between the model and the calibration. The R function `run_model` receives a vector or list of parameters to test, writes the parameters in a form that is readable for the model (e.g. via txt or csv files), runs the model (possibly via `system`) then captures and processes the model outputs. The result of the function is a “list” object with all the variables to be confronted to observed data. (C) Scheme of the calculation of the value of the objective function for a given set of parameters. For each variable, a partial value of the objective function is calculated by applying the function

specified in the column 'type' to observed and simulated values. The final value of the objective function is calculated by applying the `aggFn` to the partial function values and the weights specified in the "objective function info" table.

Table 2. Main arguments of the `calibrate` function.

| Argument | Description |
|---|---|
| `par` | A numeric vector or a list containing numeric vectors. The length of the `par` argument defines the number of parameters to be estimated (i.e. the dimension of the problem). |
| `fn` | The function to be minimized |
| `gr` | A function specifying the gradient of `fn` |
| `method` | The optimisation method to be used, 18 methods are currently available see Table 3. Different methods are allowed for each calibration phase, see 'phases' argument. |
| `upper` | Upper bound value(s) for parameters. One value or a vector of the same length as `par`. If one value is provided, it is used for all parameters. `NA` means `Inf`. By default, `Inf` is used (unconstrained). |
| `lower` | Lower bound value(s) for parameters. One value or a vector of the same length as `par`. If one value is provided, it is used for all parameters. `NA` means `-Inf`. By default `-Inf` is used (unconstrained). |
| `phases` | An optional vector of the same length as `par`, indicating the phase at which each parameter becomes active for the optimisation. If omitted, default value is 1 for all parameters. Negative integers and `NA` are accepted for phases, both meaning that the parameter will never be active, so it will remain constant throughout the calibration. |
| `control` | Fine control of the optimisation, see function help for details and Table 3. |
| `hessian` | Logical value. Should a numerically differentiated Hessian matrix be returned? |
| `replicates` | The number of replicates of model run to evaluate the objective function in case of stochastic models. One value or a vector of length `max(phases)`, to specify a different number of replicates for each phase. The default value is 1. |
| `parallel` | Logical. Use parallel computation? (e.g. for numerical gradients) |

Additionally, the user needs to specify information on the parameters to calibrate. Lower and upper bounds can be provided for each parameter for box constrained optimisation, and this often improves the parameter estimation (Bolker et al. 2013) but unconstrained optimisation is also supported. In case of a multiple phase calibration, the user must indicate the phase of the calibration where the estimation of a parameter must be included. The implementation of multiple phases in the calibration is detailed in the next section.

The `calibrar` package allows to perform the calibration with a total of 18 different optimisation algorithms allowing box constraints (Table 3), including the ones available in the package `stats` (R Core Team 2023) and `optimx` (Nash and Varadhan 2011), among others (see Table 3 for details). The calibration in multiple phases is available for all the optimisation methods through a sequential update of the objective function for each phase, built-in within the `calibrar` package. Different optimisation methods can be used for each estimation phase, allowing the combination of different methods during the course of a multi-phase calibration. The default algorithm for deterministic optimisation problems is a version of the L-BFGS-B algorithm implemented in the package `Rvmmin` (Nash et al. 2013). For stochastic optimisation problems, the default algorithm is the AHR-ES (Adaptative Hierarchical Recombination Evolutionary Strategy) which designates a novel evolutionary algorithm included in this package (Appendix 1). Another focus of this package is the handling of computationally intensive objective functions (e.g. > 60 seconds for evaluation time), including the possibility to resume interrupted optimisation runs from the intermediate "restart" files, pointing out that computation time does not need to translate into optimisation complexity. This feature is provided for the two algorithms used by default (L-BFGS-B and AHR-ES) and for the Hooke-Jeeves derivative-free minimization algorithm (HJK), a pattern search type algorithm similar to the Nelder-Mead algorithm used by default in `optim()`. Additionally, parallel computation of numerical derivatives has been implemented for all derivative-based optimisation methods, greatly improving the time needed for the optimisation.

Table 3. Optimisation methods available within the `calibrar` package.

| Algorithm | Package | Reference | Method | Notes |
|---|---|---|---|---|
| **I. Derivative-based (local) methods** | | | | |
| L-BFGS-B 3.0 | lbfgsb3c | Fidler et al. 2020 | ‘lbfgsb3’ | Default for deterministic objective functions. R interface to L-BFGS-B.3.0 (Morales and Nocedal 2011) Fortran library. |
| Algorithm 21 | Rvmmin | Nash 2013b | ‘Rvmmin’ | Variable metric method with box constraints. |
| Algorithm 22 | Rcgmin | Nash 2013a | ‘Rcgmin’ | Conjugate gradient method with box constraints. |
| spg | BB | Varadhan and Gilbert 2009 | ‘spg’ | Spectral projected gradient method. |
| nlminb | stats | R Core Team 2016 | ‘nlminb’ | Quasi-newton method with box constraints, R implementation of PORT routines (Fox 1997) |
| L-BFGS-B | stats | R Core Team 2016 | ‘L-BFGS-B’ | Original implementation in the stats package. |
| **II. Derivative-free (local) methods** | | | | |
| nmkb | dfoptim | Varadhan and Borchers 2011 | ‘nmkb’ | Nelder-Mead algorithm with box constraints. |
| hjkb | dfoptim | Varadhan and Borchers 2011 | ‘hjkb’ | Hooke-Jeeves derivative-free minimization algorithm. |
| mads | dfoptim | Varadhan and Borchers 2011 | ‘mads’ | Mesh Adaptive Direct Searches (MADS) algorithm for derivative-free and black-box optimisation |
| hjn | optimx | Nash 2016 | ‘hjn’ | Hooke and Jeeves Pattern Search Optimisation. |
| BOBYQA | minqa | Bates et al. 2022 | ‘bobyqa’ | Implementation of the BOBYQA algorithm (Powell 2009) trusted-region method. |
| **III. Heuristic (global) methods** | | | | |
| AHR-ES | calibrar | This paper | ‘AHR-ES’ | Default for stochastic objective functions. Adaptative Hierarchical Recombination Evolutionary Strategy (Appendix 1) |
| CMA-ES | cmaes | Trautmann et al. 2011 | ‘CMA-ES’ | Covariance Matrix Adaptation Evolutionary Strategy (Hansen and Ostermeier 2001). |
| SANN | stats | R Core Team 2016 | ‘SANN’ | Simulated Annealing implemented in stats::optim (method=’SANN’). |
| genSA | genSA | Xiang et al. 2012 | ‘genSA’ | Generalized Simulated Annealing. |
| DE | DEoptim | Ardia et al. 2015 | ‘DE’ | Differential evolution. |
| soma | soma | Clayden 2015 | ‘soma’ | Implementation of the Self-organizing Migrating Algorithm (Zelinka 2011). |
| genoud | rgenoud | Mebane and Sekhon 2011 | ‘genoud’ | Genetic Optimisation Using Derivatives (Sekhon and Mebane 1998). |
| pso | psoptim | Bendtsen 2012 | ‘PSO’<br>‘PSO2007’<br>‘PSO2011’ | Particle swarm optimisation (PSO), it includes two algorithms, ‘PSO2007’ and ‘PSO2011’. Using method=’PSO’ will use the default ‘PSO2007’ (see package help for details) but both can be specified. |
| hybridPSO | psoptim | Bendtsen 2012 | ‘hybridPSO’ | The method=’hybridPSO’ will use the hybrid method combining PSO and BFGS implemented in the pso package. |

## Linking to models and data

The main purpose of the package is to fit complex models to data. In order to solve a calibration problem, we first need to define the objective function. A non-intrusive black-box optimisation approach is adopted, which means that the computer code of the model to be calibrated does not require to be modified, and the model can be evaluated for a given set of parameters. For this purpose, users need to write an R function to i) write a set of parameters in the format the model is able to read, ii) run the model with this set of parameters and iii) read the model outputs back into R (Figure 1B). The output of this `run_model` function is expected to be a list, each element being one of the variables to calibrate. If the model is already implemented in R, the construction of this function is very simple. Additionally, R facilities to process and analyse data in different formats allow to handle model outputs independently of the language used for coding the model.

After the construction of the `run_model` function, the second step consists in providing information for the construction of the objective function. Each variable listed in the output of `run_model` needs to be documented in the objective function (e.g. providing the name of 'variable' and 'type', Figure 1C). This information should be provided as a `data.frame`, and will be used as an argument for the functions `calibration_data` and `calibration_ObjFn`. The `calibration_data` function is expected to read data from the disk, to produce a list with the same structure as the outputs of the `run_model` function. The function `calibration_ObjFn` will combine the observed data and the `run_model` function to create the objective function for the calibration problem (Figure 1C), which in turn will be the `fn` argument for the `calibrate` function.

To build the objective function, the 'type' selected for each variable is the function that will combine the observed and simulated data to produce a scalar value, measuring the fit between the model and the observations. Some negative log-likelihood functions are already implemented and proposed for common distributions (e.g. normal, lognormal, multinomial, Poisson; type `?objFn` to see the available functions). User defined functions can be used, as they accept two arguments (`obs` and `sim`) and return a scalar value. For example, to implement a calibration using the least squares method, we can write the following function to calculate the residual sum of squares:

```
RSS = function(obs, sim, ...) {
         value = sum((obs-sim)^2, na.rm=TRUE)
         return(value)
      }
```

The 'calibrate' column in the configuration of the objective function provides a flag to select the variables to be used for the calibration. The 'use_data' column indicates whether data are read from the disk. If `use_data=TRUE`, the file specified in the calibration settings will be used. If `use_data=FALSE`, the observed value is set to `NULL`, and the type function is expected to use simulated data only. The latter option can be particularly useful to set penalties in the model outputs or parameters, where no observed data are needed. Finally, the 'weight' column provides the relative weights to combine the values obtained for each variable. A more detailed illustration of this process is provided in the vignettes and demos of the package.

The `calibrate` function takes a list as a control argument, where fine control options are provided, for example for the parallelization of the optimisation (based on the `foreach` package, Revolution Analytics and Weston 2014). Before using the parallel implementation, a parallel 'cluster' should be created, which can be easily done using the `parallel` or `snow` R packages (see vignettes). This allows to fully control the configuration of the parallel runs, making the calibration work in different computer systems, from computers with multicore processors to high-performance supercomputers. Once the cluster is created, the `parallel=TRUE` argument must be included in the call, and the `ncores` control option should indicate the number of cores chosen (see Table 4 for details). Additionally, since each model run could require files to be written to the disk (which will be read by the `run_model` function after the simulation), a different folder needs to be assigned for each parameter combination that is tested by the optimisation algorithm or for the computation of numerical gradients. For this purpose, the `run` control option allows a directory to be specified where all the simulations are run (subfolders named `i0`, `i1`, …, `in-1` will be automatically created as needed). By default, no folders are created, so a path should be specified if the model needs to write files to the disk. All the parameter input files (Figure 1B) will be written in temporary folders (e.g. `run/i0`). The control option `master` allows to specify a folder, the full content of which will be copied to these temporary folders. Since the calibration of numerically intensive models can run for a long time, a 'restart' option is also available, allowing an interrupted calibration to be continued.

Table 4. Some options for the `control` argument of the function `calibrate`.

| Option | Description |
|---|---|
| `maxit` | Maximum number of iterations of the algorithm. |
| `ncores` | The number of cores available in the parallel cluster for the active session. If `parallel=TRUE`, the default is to get the number of cores of the system. |
| `run` | An optional folder path to create all the temporary files needed to run the simulations from the objective function for each parameter combination tested by the optimisation algorithms. |
| `master` | An optional folder path. All the contents of the `master` folder will be copied to the run folder, one copy in one subfolder per parameter combination tested at each iteration. |
| `REPORT` | Number of iterations after saving a new restart object, which contains all the information necessary to restart the calibration at that point. The default is `NULL`, and no restart files are created. |
| `restart.file` | Filename for the restart file to be created. |

## Three application examples

To illustrate the main functionality of the package, we estimated the parameters for a predator-prey model using the `calibrate` function. The model was defined by a system of ordinary differential equations for the abundance of prey $N$ and predator $P$:

$$\frac{dN}{dt} = rN\left(1-\frac{N}{K}\right) - \alpha NP$$
$$\frac{dP}{dt} = -lP + \gamma\alpha NP$$

The parameters to estimate were the prey's growth rate $r$, the predator's mortality rate $l$, the carrying capacity of the prey $K$ and $\alpha$ and $\gamma$ for the predation interaction. To start, we created the demonstration data for this model using the function `calibrar_demo` function with `T=100` as an additional argument to specify the time horizon.

```
LV        = calibrar_demo(path=path, model='PredatorPrey', T=100)
setup     = calibration_setup(file = LV$setup)
observed  = calibration_data(setup=setup, path=LV$path)
run_model = calibrar:::.PredatorPreyModel
obj       = calibration_objFn(model=run_model, setup=setup, observed=observed, T=LV$T,
                              aggregate=TRUE)
```

To run the calibration, we needed to specify the initial guess for the parameter values (`par`), the objective function to minimize (`fn`) and optionally the lower and upper thresholds for the parameters (`lower` and `upper`) and the phase number at which each parameter needs to be estimated (`phases`).

```
calibrate(par=LV$guess, fn=obj, lower=LV$lower, upper=LV$upper, phases=LV$phase)
```

The argument `method` can be used to change the default optimisation algorithm. We calibrated the model using five different optimisation algorithms with their default parameters: i) derivative-based: Algorithm 21 (L-BFGS-B, default for deterministic functions, from package `Rvmmin`), the conjugate gradient method (CG from the `Rcgmin` package); ii) derivative free: Nelder-Mead (default R optimiser, `stats::optim()`); iii) heuristic: AHR-ES (default for stochastic functions), and the CMA-ES (from `cmaes` package). The results are shown in Figure 2 and Table 5, where we can observe that the algorithm Nelder-Mead and CG could not find the solution using the default parameters provided in the original optimisation functions (while some improvements may be obtained after some tuning). The full code can be found in the vignettes of the package.

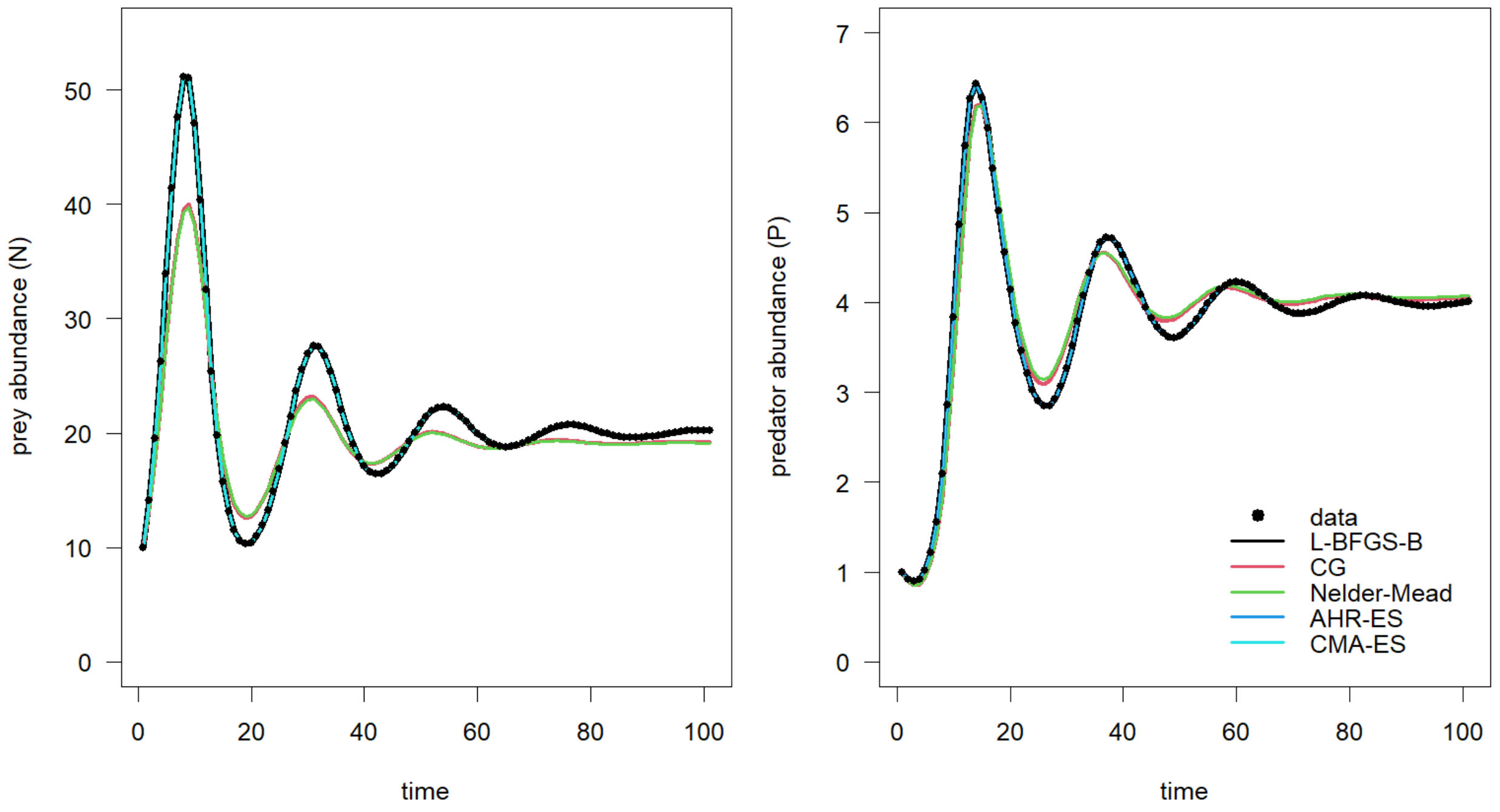

Figure 2. Results of the calibration of the predator-prey model using different optimisation methods. The simulated data (points) and model fits (lines) are shown. For the optimisation methods “AHR-ES”, “L-BFGS-B” and “CMA-ES” there are no visual differences and the lines merge.

Table 5. Summary of the calibration results for the predator-prey model using five different optimisation methods with default parameters.

| **Method** | **Objective function value** | **Parameter** | | | | |
|---|---|---|---|---|---|---|
| | | **R** | **l** | **K** | **α** | **γ** |
| data | 4.96E-07 | 0.5 | 0.2 | 100 | 0.1 | 0.1 |
| L-BFGS-B (Rvmmin) | 5.61E-07 | 0.4999 | 0.2000 | 99.9960 | 0.1000 | 0.1000 |
| CG (Rcgmin) | 1.29961642 | 0.4489 | 0.2746 | 67.4768 | 0.0796 | 0.1802 |
| Nelder-Mead | 1.41516908 | 0.4500 | 0.2751 | 66.2434 | 0.0788 | 0.1827 |
| AHR-ES | 5.61E-07 | 0.4999 | 0.2000 | 99.9960 | 0.1000 | 0.1000 |
| CMA-ES | 5.61E-07 | 0.4999 | 0.2000 | 99.9960 | 0.1000 | 0.1000 |

A second example involves the calibration of a Poisson Autoregressive Mixed model for the dynamics of a population in different sites:

$$\log(\mu_{i,t+1}) = \log(\mu_{i,t}) + \alpha + \beta X_{i,t} + \gamma_t,$$

where $\mu_{i,t}$ is the size of the population in site $i$ at year $t$, $X_{i,t}$ is the value of an environmental variable in site $i$ at year $t$. The parameters to estimate were $\alpha$, $\beta$, and $\gamma_t$, the random effects for each year ($\gamma_t \sim N(0, \sigma^2)$), and the initial population at each site $\mu_{i,0}$. We assumed that the observations $N_{i,t}$ follow a Poisson distribution with mean $\mu_{i,t}$. We could also create the data for this model using the function `calibrar_demo`, with the additional arguments `L=5` (five sites) and `T=100` (one hundred years):

```
ARPM    = calibrar_demo(path=path, model="PoissonMixedModel", L=5, T=100)
```

```
setup    = calibration_setup(file=ARPM$setup)
observed = calibration_data(setup=setup, path=ARPM$path)
forcing  = as.matrix(read.csv(file.path(ARPM$path, "master", "environment.csv"),
                              row.names=1))
```

For this example, the `run_model` function includes a penalty for the time series of $\gamma_t$ that was needed to construct the objective function:

```
run_model = function(par, forcing) {
  output = calibrar:::.PoissonMixedModel(par=par, forcing=forcing)
  output = c(output, list(gammas=par$gamma))
  return(output)
}

obj  = calibration_ObjFn(run_model=run_model, info=calibrationInfo, observed=observed,
                         forcing=forcing, aggregate=TRUE)
```

The calibration was run as in the previous example calling the `calibrate` function:

```
calibrate(par=ARPM$guess, fn=obj, lower=ARPM$lower, upper=ARPM$upper, phases=ARPM$phase)
```

Here we calibrated the model using four different optimisation algorithms: i) derivative-based: L-BFGS-B (`stats`) and L-BFGS-B v3 , ii) derivative free: HJKB; iii) heuristic: AHR-ES and SANN. The results are shown in Figure 3 and Table 6. In this case, the SANN algorithm could not find the solution despite an increased number of maximum iterations. The best solution found was using HJKB algorithm, closely followed by the AHR-ES and L-BFGS-B algorithms (Table 6 and Figure 3). The full code can be found in the vignettes of the package.

Table 6. Summary of the calibration results for the autoregressive Poisson mixed model using five different optimisation methods.

| **Method** | **Elapsed time (seconds)** | **Objective function value** | **Parameters** | | | | | |
|---|---|---|---|---|---|---|---|---|
| | | | **α** | **β** | **$\gamma_t$** | | | |
| | | | | | **mean** | **s.d.** | **correlation** | **bias** |
| data | - | -186178.145 | 0.4000 | -0.4000 | -0.0187 | 0.1952 | 1.000 | 0.0000 |
| L-BFGS-B | 57 | -186081.484 | 0.4846 | -0.4157 | -0.0867 | 0.1923 | 0.904 | 0.0680 |
| L-BFGS-B 3.0 | 14 | -180576.287 | 0.3700 | -0.1389 | -0.2528 | 0.0460 | 0.311 | 0.2341 |
| HJKB | 222 | -186094.819 | 0.3923 | -0.4145 | 0.0039 | 0.1869 | 0.908 | -0.0226 |
| AHR-ES | 612 | -186085.887 | 0.3967 | -0.4143 | -0.0009 | 0.1940 | 0.862 | -0.0178 |
| SANN | 43004 | 1.5117E+16 | 0.2000 | 0.1000 | 0.0000 | 0.0000 | - | -0.0187 |

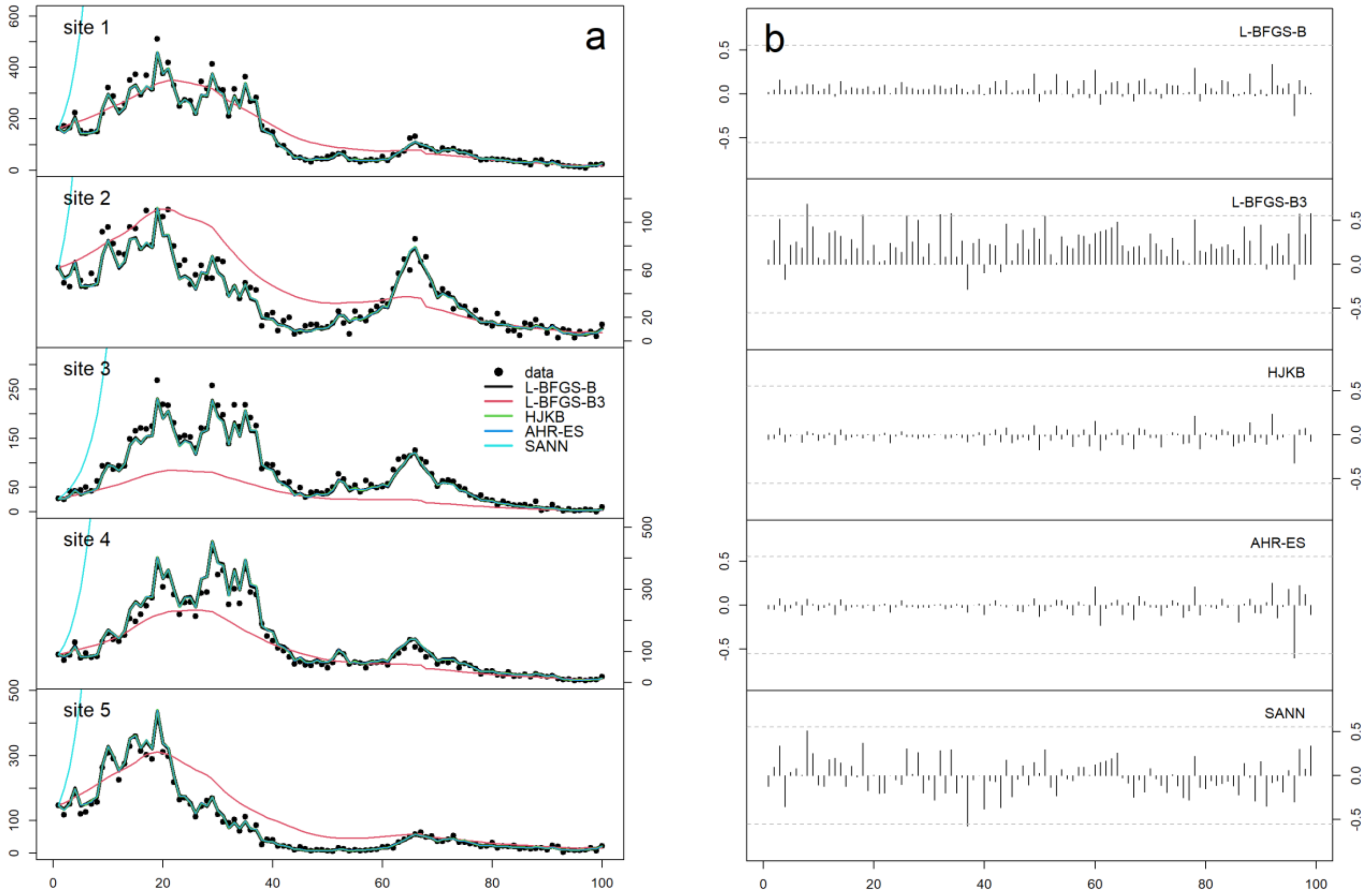

Figure 3. Results of the calibration of the autoregressive Poisson mixed model using different optimisation methods. (a) The simulated data (points) and model fits (lines) are shown. For the optimisation methods "HJKB" , "AHR-ES" and "L-BFGS-B" there are no visual differences and the lines merge. (b) Time series of differences between the $\gamma_t$ estimated by each algorithm and the real parameter values. For each case, the dotted lines represent the 95% limits for the differences

A third and more complex application involved the calibration of the stochastic individual-based model OSMOSE, a multispecies spatially-explicit ecosystem model (Shin and Cury 2001, 2004). It has been applied in the Northern Peru Current Ecosystem (Oliveros-Ramos et al. 2017) to model the life history and spatiotemporal dynamics of nine interacting species (one macro-zooplankton group, one crustacean, one cephalopod and six fish species), between 1992 and 2008, from 20ºS to 6ºN and 93ºW to 70ºW, with 1/6º of spatial resolution. The model was confronted to time series of abundance indices, fisheries landings and catch-at-length data. The objective function used a penalized likelihood approach, combining log-normal and multinomial likelihoods. A total of 307 parameters were estimated in 4 sequential phases. The OSMOSE model is implemented in the Java language, and it was not an option to recode it in another language for parameter estimation purposes. Each calibration trial lasted 5 days using a High-Performance Computing (HPC) cluster under the Portable Batch System (PBS) for jobs scheduling, used 64 cores and needed to be relaunched every 24 hours due to system restrictions. The objective function for this model is stochastic and was optimized using the AHR-ES algorithm. The calibration in sequential phases improved the final parameter estimates as described in Oliveros-Ramos et al. (2017). The calibrar package greatly simplified the task of the calibration of a complex model like OSMOSE, which otherwise can be highly technical, if not impossible, with other optimisation packages. The scripts used for this calibration which may be adapted for applications of the calibrar package in other HPC systems are available at `https://github.com/osmose-model/calibrar.osmose`. For other examples, the calibrar package has been used in the model fitting of some complex models like in Oliveros-Ramos et al. (2010), Oliveros-Ramos and Peña (2011), Grüss et al. (2015), Grüss et al. (2016), Dueri et al. (2016), Halouani et al. (2016), Travers et al. (2019), Moullec et al. (2019), Banaru et al. (2019), Xing et al. (2022) and Morell et al. (2023).

## Comparison with other software

Implementation of general-purpose optimisers can be found in R (see Optimisation and Mathematical Programming Task View at CRAN: http://cran.r-project.org/web/views/Optimisation.html). Two very useful features for model calibration are the performance of constrained optimisation (limiting the search to a box by defining lower and upper boundaries to parameter values) and the calibration in multiple phases (to improve the search of the global minimum by performing a sequential approximation). The former option is implemented in several R packages, including the `optim` function (providing the "L-BFGS-B" method, Byrd et al. 1995) and several others wrapped in the `optimx` package (Nash and Varadhan 2013). The latter option is available in some other R packages (e.g. `Rcgmin` and `Rvmmin`) for a single optimisation, but a sequential calibration, as described here, would have to be performed manually. Additionally, the calibrar package allows to choose different optimisation methods for each calibration phase, giving the possibility to mix heuristic global optimisation methods and local optimisation ones to improve the performance of a multiple phase calibration.

For the particular purpose of the calibration of stochastic models, several meta-heuristic and non-derivative based algorithms are now available in R, from EAs (e.g. `genalg`, `DEoptim` and `cmaes` packages) to other nature-inspired algorithms (e.g. Simulated Annealing '`SANN`' method in `optim` and the Particle Swarm Optimisation (PSO) algorithm in the `hydroPSO` package, Zambrano-Bigiarini and Rojas 2013). However, while all of these packages and algorithms provide support for constrained optimisation, none of them provides support for keeping fixed parameters during the course of a single optimisation, and a multiple phase calibration would have to be performed manually by modifying the objective function for each trial. Furthermore, from the implementation point of view, a very important feature for the calibration of complex models is the parallel implementation of the optimisation routine. The PSO algorithm in the `hydroPSO` has its parallel implementation tied to the core of the function and does not allow its use in high-performance clusters, especially under a queue system, and only the `DEoptim` package provides a more flexible externally configured parallelization.

Additionally, in the construction of the objective function, calibrar allows an easy transferability of the calibration problem to other general-purpose optimizers, which can be useful under certain circumstances (e.g. see Bolker et al. 2013). There is indeed "no free lunch" in optimisation, and no optimisation algorithm will perform better than all others for every type of optimisation problems (Wolpert and Macready 1997). Other calibration-oriented packages like `hydroPSO` provide functions to write parameters and read outputs, but this approach breaks the "objective function" approach for the optimisation, and while the `hydromad` package (Andrews et al. 2011) offers support for the automated construction of an objective function in a standard way, it is restricted to some particular cases useful in hydrological modelling. In these regards, the package `calibrar` is meant to be generic enough to be used in a variety of optimisation problems, including the calibration of complex (i.e. non-linear, with a lot of parameters and long runtime) and stochastic models. Three features of `calibrar` render it particularly useful for the calibration of computationally intensive stochastic models: the parallelization of the simulations, the ability to handle replicate simulations in the evaluation of the objective function and the 'restart' option, which allows the calibration of complex models to be handled under restricted access to high performance resources (e.g. clusters with queue systems and fixed wall time).

## Conclusions and perspectives

A successful model calibration implies several computational, theoretical and practical challenges. The `calibrar` package intends to provide a framework to simplify the calibration of models, in particular complex and stochastic ones (e.g. Individual-Based Models), for which fewer developments have been done compared to those for deterministic and differentiable models. To our knowledge, the `calibrar` package is the first one wrapping several global search optimisation methods, facilitating their use and testing for parameter estimation. We adopted a "black-box" and "non-intrusive" approach, since most complex models are often computationally intensive and most likely implemented in fast low-level languages and recoding for calibration purposes is not the best option. The restart functionality can help in the optimisation of computationally intensive problems, since most of the computing time is spent evaluating the objective function (Nash 2014) and the ability to restart an interrupted optimisation could save significant amounts of time. The use and testing of the `calibrar` package with several real-world optimisation problems for the calibration of complex ecological models has directed the developments of the package and its current flexibility, while we expect future applications will help to continue improving the package given its open-source nature.

## Acknowledgements

This research has been funded by the Biodiversa and Belmont Forum project SOMBEE (BiodivScen ERA-Net COFUND programme, ANR contract n°ANR-18-EBI4-0003-01), the European Union's Horizon 2020 research and innovation programme under grant agreements No 869300 (FutureMARES), No 817578 (TRIATLAS), the Horizon Europe research and innovation programme under grant agreement No 101060072 (ActNow), France Filière Pêches, the Pew marine fellows programme, and the Fondation pour la Recherche sur la Biodiversité (EMIBIOS, FRB contract no. APP-SCEN-2010-II). ROR was supported by an individual doctoral research grant (BSTD) from the "Support and training of scientific communities of the South" Department of IRD, managed by Campus France. We thank David Kaplan, Coleen Moloney, and Arnaud Bertrand for their advice during the development of this research. We acknowledge Philippe Verley, Nicolas Barrier, Morgane Travers-Trolet, Laure Velez, Criscely Luján and Alaia Morell for testing and helping to improve the `calibrar` package. The authors would like to thank the Pôle de Calcul et de Données Marines (PCDM) for providing access to their DATARMOR computing resources (https://wwz.ifremer.fr/pcdm/Equipement). This work is a contribution from the Cooperation agreement between the Instituto del Mar del Peru (IMARPE), the University of Cape Town (UCT) and the Institut de Recherche pour le Développement (IRD), through the LMIs DISCOH and ICEMASA.